# Geometric-phase microscopy for high-resolution quantitative phase imaging of plasmonic metasurfaces with sensitivity down to a single nanoantenna

Petr Bouchal,[1,2,*] Petr Dvořák,[1,2] Jiří Babocký,[1,2] Zdeněk Bouchal,[3] Filip Ligmajer,[1,2] Martin Hrtoň,[1,2] Vlastimil Křápek,[1,2] Alexander Faßbender,[4] Stefan Linden,[4] Radim Chmelík,[1,2] Tomáš Šikola[1,2]

[1]*Institute of Physical Engineering, Faculty of Mechanical Engineering, Brno University of Technology, Technická 2, 616 69 Brno, Czech Republic*
[2]*Central European Institute of Technology, Brno University of Technology, Purkyňova 656/123, 612 00 Brno, Czech Republic*
[3]*Department of Optics, Palacký University, 17. listopadu 1192/12, 771 46 Olomouc, Czech Republic*
[4]*Physikalisches Institut, Universität Bonn, Nussallee 12, 53115 Bonn, Germany*
*\*petr.bouchal@ceitec.vutbr.cz*

**Optical metasurfaces have emerged as a new generation of building blocks for multi-functional optics. Design and realization of metasurface elements place ever-increasing demands on accurate assessment of phase alterations introduced by complex nanoantenna arrays, a process referred to as quantitative phase imaging. Despite considerable effort, the widefield (non-scanning) phase imaging that would approach resolution limits of optical microscopy and indicate the response of a single nanoantenna still remains a challenge. Here, we report on a new strategy in incoherent holographic imaging of metasurfaces, in which unprecedented spatial resolution and light sensitivity are achieved by taking full advantage of the polarization selective control of light through the geometric (Pancharatnam-Berry) phase. The measurement is carried out in an inherently stable common-path setup composed of a standard optical microscope and an add-on imaging module. Phase information is acquired from the mutual coherence function attainable in records created in broadband spatially incoherent light by the self-interference of scattered and leakage light coming from the metasurface. In calibration measurements, the phase was mapped with the precision and spatial background noise better than 0.01 rad and 0.05 rad, respectively. The imaging excels at the high spatial resolution that was demonstrated experimentally by the precise amplitude and phase restoration of vortex metalenses and a metasurface grating with 833 lines/mm. Thanks to superior light sensitivity of the method, we demonstrated, for the first time to our knowledge, the widefield measurement of the phase altered by a single nanoantenna, while maintaining the precision well below 0.15 rad.**

**Introduction**

Shaping of light in photonics has traditionally been governed by natural optical phenomena manifested by refraction, diffraction and birefringence in conventional optical media. Nowadays, ever increasing attention is attracted by composite artificial nanostructures that are collectively referred to as metasurfaces.[1,2] The metasurfaces are made of structured arrays of subwavelength-spaced nanoantennas that can provide resonant enhancement of electromagnetic waves with spatially controlled phase variations.[3] Compared to traditional refractive optics, metasurface-based components are physically thin and introduce abrupt phase changes observable in the far field. This makes them promising candidates for a new generation of multi-functional optics[4,5] with unique properties for imaging,[6–8] biosensing,[9,10] beam steering,[11,12] shaping of special states of light,[13–16] holography,[17–20] adaptive light modulation[21–23], and nonlinear optics.[24,25] For studies on metasurfaces and control of their fabrication and light-shaping capabilities, quantitative information about the phase imposed on the light scattered by individual nanoantennas becomes of great importance. Since information acquired from the far-field distribution of light is not sufficient, the high-resolution quantitative imaging of the phase directly in the metasurface plane is highly appreciated.

During the past years, this task has been addressed only in a few approaches including scanning near-field optical microscopy and interference-based techniques. Scattering-type scanning near-field optical microscopy (s-SNOM) provides subdiffraction resolution and allows imaging of complex nanoantenna arrays including their near-field phase response.[26,27] The application of s-SNOM is limited by time-consuming measurements typical for inherently slow scanning techniques and its use is preferably aimed at infrared region. Most phase measurements were realized using interference-based techniques working either with broadband continuous light sources[28–31] or ultrashort laser pulses.[32–35] The potential of these methods is reduced by the necessity of dispersion compensation and scanning of the sample to acquire widefield information. These drawbacks are partly solved in coherence controlled holographic microscopy.[36] This technique provides achromatic and widefield imaging, but its implementation requires a very complex and vibration sensitive experimental setup and troublesome filtering of the leakage light coming from the driving field. Design and demanding adjustment of the system do not allow the use of high-aperture microscope objectives; hence the spatial resolution cannot reach the full potential of optical microscopy.[36] Although the diffraction limit attainable in optical imaging is close to the size of nanoantennas, widefield imaging of metasurfaces with the resolution and the sensitivity sufficient for phase monitoring of a single nanoantenna has not yet been demonstrated and still remains a challenge.

In this article, we present a new experimental strategy that overcomes the current state of the art by providing unprecedented spatial resolution and light sensitivity in the widefield quantitative mapping of plasmonic metasurfaces. The measurement is based on the incoherent correlation holography enhanced by the polarization-selective control of light. This technique utilizes optical elements for transformation of the geometric phase, referred to as fourth-generation (4G) optics.[37–39] Although here demonstrated for Pancharatnam-Berry phase metasurfaces working with circularly polarized light, the polarization multiplexing implemented by 4G optics makes the method inherently predestined for the restoration of phase differences introduced between any orthogonal polarization components. The phase is quantitatively reconstructed from the mutual coherence function provided by off-axis holographic correlation records. The metasurfaces are excited by broadband spatially incoherent illumination and unlike other methods, the leakage light is not canceled, but its partial spatial correlation with the scattered light is used to capture the holographic records. In experiments carried out in an inherently stable common-path arrangement, we show that the developed quantitative phase measurement is a powerful tool for single-shot widefield inspection of metasurfaces, meeting the highest demands on precision and accuracy. All these attributes are expected to be essential for an efficient quantitative analysis of the increasingly more complex metasurfaces, among them certainly being those designed to manipulate the orbital angular momentum of light (OAM).[40,41] We demonstrate the abilities of our experimental setup on two members of this class of metasurfaces, namely, we present here high-resolution phase images of vortex metalenses with different topological charges[14,15,16] and of a metasurface mask for the generation of vortex Laguerre-Gaussian (L-G) beam.[42,43] The exceptional sensitivity of the method is further evidenced by the first scanning-free measurement capable of quantification of the phase altered by a single nanoantenna. The measurements of an isolated nanostructure enable the assessment of effects in its phase response that would be otherwise obscured by its surroundings and therefore facilitate studies on correlation between morphology, material properties and optical properties on the lowest level achievable by optical microscopy.[44,45]

**Results**

*Experimental strategy*

The developed phase imaging technique, in this article referred to as quantitative 4G optics microscopy (Q4GOM), was implemented in Nikon Eclipse L150 microscope working with an add-on geometric-phase imaging module. This module benefits from the unique capabilities of a geometric-phase grating (GPG), providing polarization-selective angular separation of incident waves (for more details see Materials and Methods). The setup used for the phase measurement of the metasurfaces is shown in Figure 1a and schematically illustrated in Figure 1b.

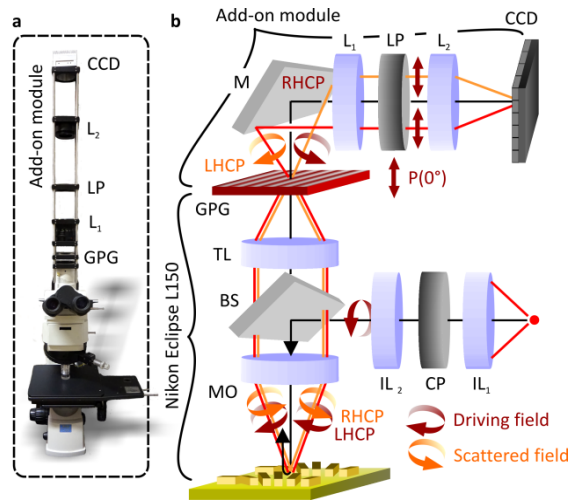

**Figure 1** Experimental setup for high-resolution, widefield measurement of phase alterations introduced by plasmonic metasurfaces: RHCP–right-handed circular polarization, LHCP–left-handed circular polarization, MO–microscope objective,

BS–beam splitter, TL–tube lens, GPG–geometric phase grating, M–mirror, L1–first Fourier lens, LP–linear polarizer, L2–second Fourier lens, CCD–charged coupled device.

In the measurements, the microscope was operated with a standard illumination system composed of a tungsten-halogen lamp (filtered at central wavelength $\lambda_C$ = 600 nm, full width at half maximum $\Delta\lambda$ = 50 nm), lenses $IL_1$, $IL_2$ providing Köhler illumination conditions and a circular polarizer CP used for excitation of metasurfaces by light with left-handed circular polarization (LHCP). The sample is illuminated through a beam splitter BS and a microscope objective MO (Nikon 10x, NA = 0.3 or Nikon 100x, NA = 0.9) acting simultaneously as a condenser and an imaging lens. In the case of Pancharatnam-Berry phase metasurfaces, the light coming from the metasurface is composed of a leakage wave originating from the driving light and maintaining its LHCP, and the scattered wave that holds right-handed circular polarization (RHCP). The polarization-coded waves are directed through the BS and a tube lens TL (Nikon CFI60) toward GPG placed in the back focal plane of TL. The polarization state of the waves impinging on the GPG is changed from LHCP to RHCP, and vice versa, while a geometric phase of opposite sign is imposed on their wavefronts.[37–39] This way, the leakage and scattered waves are angularly separated and laterally sheared at the back focal plane of the first Fourier lens $L_1$. To achieve interference of the leakage and scattered waves at the image plane of a CCD, their circular polarizations are projected to a linear polarization by the polarizer LP. Subsequently, the waves are recombined by the second Fourier lens $L_2$ creating an off-axis hologram recorded with an achromatic carrier frequency.

*Calibration measurements*

The calibration of Q4GOM can hardly be done using metasurfaces, since measurement inaccuracies will be combined with fabrication imperfections. An alternative measurement providing the ground-truth phase values is available through the polarization-division multiplexing. This strategy makes Q4GOM predestined to measure differences in both geometric and dynamic phase introduced between any orthogonal polarization components (the geometric phase depends on the path in the parameter space and varies when the polarization of light is transformed, while the dynamic phase is related to the speed of light, i.e. the refractive index). The measurement of the dynamic phase introduced between orthogonal linear polarizations by optical birefringence is used for the calibration of the setup. In the calibration measurement, a spatial light modulator (SLM, Hamamatsu X10468-01) is deployed as a birefringent device and its active surface is placed at the front focal plane of the microscope objective. By operation of the SLM, the dynamic phase of the linearly polarized light is electronically controlled, while maintaining the phase of the orthogonal linear polarization unchanged.[46,47] This way, the ground-truth phase values can easily be set. In the calibration measurement, the SLM was addressed by a four-quadrant driving map providing the phase values 0, $\pi/2$, $\pi$, $3\pi/2$ in the individual quadrants (inset in Figure 2a).

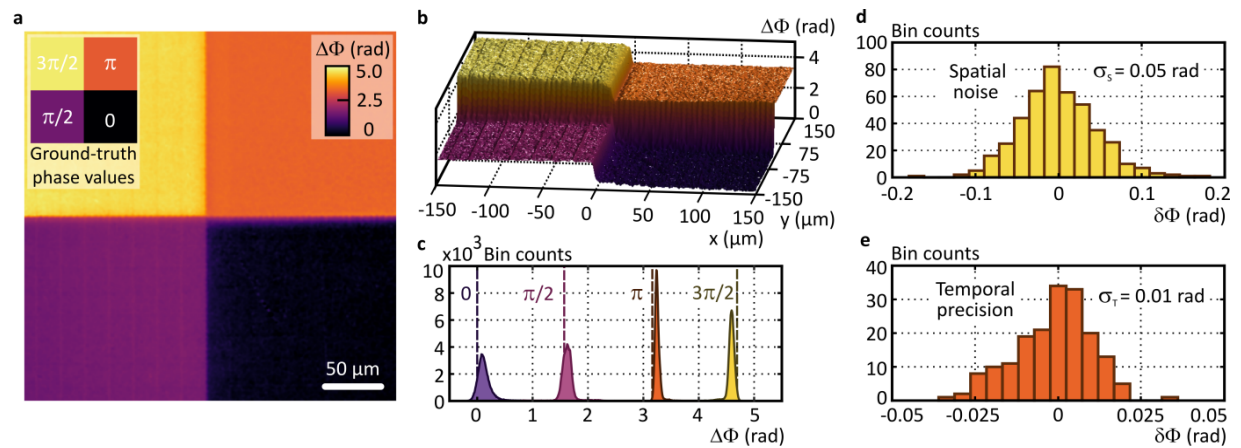

**Figure 2** Evaluation of the ground-truth accuracy and the spatial and temporal precision in the calibration phase measurement. (a) Driving phase mask displayed on the SLM (inset top left) and the color-coded phase acquired in our setup. (b) Three-dimensional visualization of the restored phase from (a). (c) Histogram of the phase restored in individual quadrants in (a). (d) Histogram of the background phase noise evaluated in area 10x10 µm². (e) Histogram of temporal phase fluctuations during 15-minute-long period.

The correlation records were taken with the microscope objective 10×, NA = 0.3, and four different phase levels set by the SLM were reconstructed. The color-coded restored phase $\Delta\Phi$ and its three-dimensional visualization are shown in Figure 2a,b. To evaluate the accuracy of the method, the histograms of the restored phase were created and their calculated mean values compared with the ground-truth phase values set by the SLM (Figure 2c). In all quadrants, the deviations were less than 0.1 rad, which corresponds to the optical path difference $\lambda_C/60$. When evaluating the spatial background noise in the area $(10\times10)$ µm², the variance $\sigma_S$ = 0.05 rad was obtained (Figure 2d). The temporal precision determined from 180 correlation records taken during a 15-minute-long period resulted in a variance of $\sigma_T$ = 0.01 rad (Figure 2e). The phase fluctuations were evaluated in the area corresponding to the diffraction spot of the microscope objective used in the experiment. The calibration measurements were carried out in a common optical laboratory without using a covering box or any vibration protection.

*Measurement of nanoantenna arrays*

To demonstrate the capabilities of Q4GOM for the inspection of metasurfaces, a benchmark sample using the Pancharatnam-Berry phase and providing gradual variations of the geometric phase across the range $\langle 0,2\pi \rangle$ was prepared. The sample is composed of identical nanoantennas schematically illustrated in Figure 3a. The directions of their axes are spatially varying as shown by the image from the scanning electron microscope (SEM) in Figure 3b. Setting the nanoantenna angle to φ causes a local change in the geometric phase of 2φ.[48] The phase is altered by the change of handedness of the circular polarization, in which LHCP of driving light is transformed to RHCP of scattered light. The benchmark sample used in the measurement is composed of 19 square areas with the size of $(10\times10)$ µm². In the areas numbered in ascending order, the angular rotation of the nanoantennas increases with a step of 10°. Visualization of nanoantennas in the individual areas is provided by SEM images shown in Figure 3c.

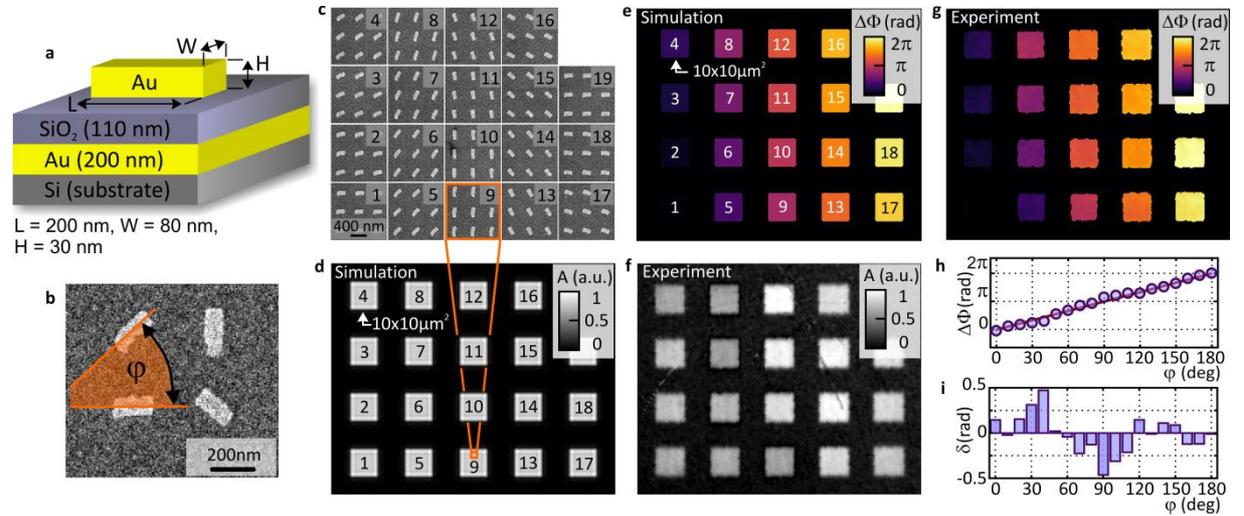

**Figure 3** Phase imaging of the benchmark sample. (a) Schematic illustration of a single nanoantenna. (b) SEM image of the nanoantennas with varying angular orientation. (c) Enlarged parts of SEM images showing the orientation of the nanoantennas in individual areas of the benchmark test. (d) Simulation of the amplitude of light scattered by the benchmark sample. (e) Simulation of the phase response of the benchmark sample. (f) Amplitude and (g) phase image of the benchmark test restored from correlation records acquired by Q4GOM. (h) Theoretical dependence of the geometric phase on the rotation angle of nanoantennas (solid line) and values of the phase measured in individual square areas of the benchmark test (circles). (i) Differences between theoretical dependence and values of the phase measured in individual square areas of the benchmark test.

To compare theory and experiment, the distributions of both amplitude and phase of the scattered light in the plane of benchmark metasurface were simulated by finite difference time domain method (FDTD) combined with Green's function formalism (for more details see Materials and Methods). The gray-scale amplitude image and color-coded phase image are presented in Figure 3d,e. Subsequently, the benchmark sample was recorded using the MO 10×, NA = 0.3 and both amplitude and phase of the sample were reconstructed. The experimentally restored amplitude image is shown in Figure 3f, while the phase image of the nanoantenna array is illustrated in Figure 3g. During the processing of experimental data, the background noise in the phase image was suppressed by a binary mask created from the measured amplitude. In Supplementary information, a raw phase image is shown and the used masking procedure described. The compliance between the simulation and the experimental results is demonstrated in Figure 3h,i. The solid line in Figure 3h illustrates the theoretical dependence of the geometric phase on the angular orientation of the nanoantennas that was obtained from simulation of their phase response. The circles represent the experimental values of the phase restored in the individual square areas of the benchmark sample. Differences between theoretically and experimentally obtained phase values are presented in Figure 3i. The greatest deviations of the measured phase from the theoretical values occur for the nanoantenna angles φ = 40° and 90° and correspond to 0.5 rad. Because the spatial background noise verified by the calibration measurement is much smaller, we can conclude that the deviations have their origin in manufacturing imperfections rather than in the measurement accuracy. This confirms that Q4GOM can be deployed as a powerful tool for the inspection of optical metasurfaces, providing data important to optimize the metasurface elements and thus increase their performance.

*High-resolution measurement of metasurface phase gratings*

With the current state of the art, the widefield imaging of metasurfaces cannot be realized with high spatial resolution. In our experiments, the correlation records of a periodic metasurface structure were successfully captured and reconstructed using a high aperture MO (100×, NA = 0.9). To demonstrate the performance of our high-resolution imaging method, the restored phase images were examined in comparison with the results obtained by a common MO (10×, NA = 0.3). For the measurement, we prepared two different metasurface phase gratings composed of parallel areas with the width of 2.75 μm and 1.2 μm, respectively, in which the phase values $-\pi/2$ and $\pi/2$ alternate. In the first measurement, the phase image of the low frequency metasurface grating (364 lines/mm) was restored from the records taken with the MO 10× and subsequently with the MO 100×. The obtained color-coded phase images are shown in Figures 4a and 4b, respectively. The measurement proved that Q4GOM is capable of metasurface imaging carried out with the high aperture MO, while maintaining the ground-truth accuracy of the restored phase. This is clearly documented by a very good compliance of the cross-section profiles I and II in Figure 4c, obtained in low and high aperture measurements. The phase images obtained in the measurements carried out with a high frequency metasurface grating (833 lines/mm) using the MO 10× and the MO 100× are shown in Figures 4d and 4e, respectively. Insufficient spatial resolution of the MO 10× results in errors in the restored phase that are evident in the cross-section profile III in Figure 4f. In the measurement realized with the high aperture MO 100×, the phase of the metasurface grating is still restored correctly, as clearly demonstrated by the cross-section phase profile IV in Figure 4f. The experiments using low and high frequency metasurface gratings prove that Q4GOM allows the quantitative measurement of the phase realized with the high aperture MO, while keeping the highest demands on the ground-truth accuracy of the restored phase. Since the size and spacing of the nanoantennas are both at deeply sub-wavelength scale, the demonstrated high-resolution phase measurement is of great importance for applications based on steep spatial phase variations like metaholograms, metalenses with a short focal length or vortex masks of complex topology.

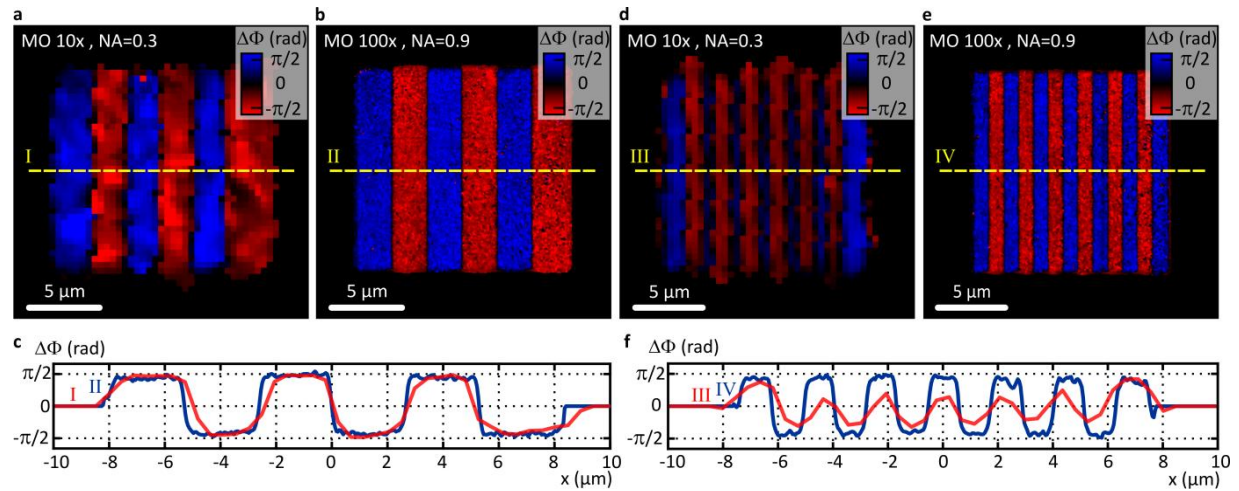

**Figure 4** Phase imaging of metasurface gratings. (a) Color-coded phase image of low frequency grating (364 lines/mm) obtained with the MO 10x, NA = 0.3. (b) The same as in (a) but for the MO 100x, NA = 0.9. (c) Cross-section phase profiles along the dashed lines I and II. (d) Color-coded phase image of high frequency grating (833 lines/mm) obtained with the MO 10x, NA = 0.3. (e) The same as in (d) but for the MO 100x, NA = 0.9. (f) Cross-section phase profiles along the dashed lines III and IV.

*High-resolution imaging of vortex metastructures*

Design and implementation of metasurface elements have experienced enormous advances in their complexity and functionality over the past decade. Here we demonstrate the successful use of the Q4GOM for inspection of such metasurface elements carried out with high spatial resolution and accuracy. In the measurements realized with the high aperture MO 100×, NA = 0.9, a metasurface mask for the generation of a vortex L-G beam and vortex metalenses of different topological charges were used as samples. The results obtained in the imaging of the mask for the generation of the L-G beam (topological charge 3, radial index 3) are shown in

Figure 5. The amplitude and phase images reconstructed from the records captured in Q4GOM setup are shown in Figures 5a,b. Magnified images in the insets show that the high spatial resolution and sensitivity of the imaging enable visualizing areas with the same orientation of the nanoantennas. This allows the tracking of local errors and the inspection of the metasurfaces at the level of their basic building blocks. Such capability is well documented by the cross-section amplitude profile in Figure 5c taken along the red line I in enlarged detail of Figure 5a. The peaks appearing periodically in the amplitude profile have the width of approximately 0.3 µm and were created by collecting light scattered by the individual nanoantennas. The phase response of the individual nanoantennas is demonstrated by the cross-section phase profile in Figure 5d taken along the blue line II in Figure 5b. Near the amplitude peaks marked by the red rings, the phase remains approximately constant, creating a stepped profile. The phase levels at the individual steps represent the phase changes caused by the relevant nanoantennas. The height of each step is approximately 0.3 rad and represents the angle of rotation between adjacent nanoantennas. The phase images of the vortex metalenses are shown in Figures 5e,f. The number of spiral arms and the fringe density represent the topological charge and the curvature of the wavefront just in the plane of the metasurface. The metalens with a longer focal length has the topological charge 1 (Figure 5e), while the two-arm spiral of the metalens with a shorter focal length indicates the topological charge 2 (Figure 5f). The high spatial resolution and sensitivity of the method are again well documented by enlarged details of the phase maps showing a pixel-like structure of the restored phase. Information on the finest phase structure provided by the measurement is valuable for fundamental studies on natural and artificial phase singularities and evaluation of fabrication errors of complex nanoantenna arrays.

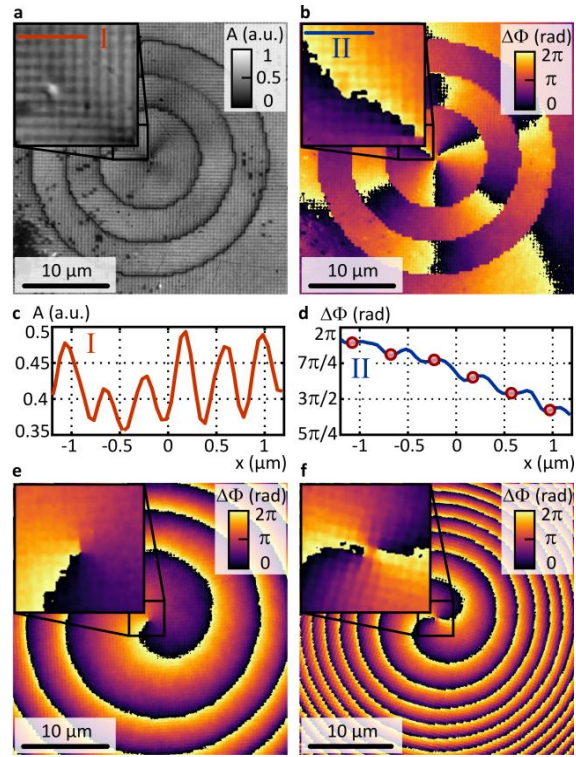

**Figure 5** High-resolution imaging of vortex metastructures by Q4GOM. (a) Amplitude and (b) phase image of a metasurface mask for generation of vortex L-G beam (topological charge 3, radial index 3). (c) Cross-section amplitude profile along red line I in (a) with the peaks distinguishing individual nanoantennas. (d) Cross-section phase profile along blue line II in (b) with the steps representing phase alteration by individual nanoantennas. (e) Phase image of a vortex metalens with topological charge 1. (f) The same as in (e) but for shorter focal length and topological charge 2. Pixel-like structure in enlarged details indicates phase of light scattered by individual nanoantennas.

*Quantitative phase measurement of a single nanoantenna*

To convincingly demonstrate the unique capability of Q4GOM to capture and quantify the geometric phase altered by a single nanoantenna, we prepared a metasurface array composed of nanoantennas with the spacing of 2.5 µm and alternating angular orientation, which introduced the geometric phase with values −π/2 and π/2, respectively. The enhanced spacing of the nanoantennas was chosen to assure an independent operation, in which restoration of the single nanoantenna phase remains unaffected by the signal from other nanoantennas. The SEM image of the sample is shown in Figure 6a. The correlation records of the sample were captured in Q4GOM setup using the high aperture MO 100×, NA = 0.9. The reconstructed amplitude image shown in Figure 6b proves that the sensitivity of Q4GOM is sufficient to collect light scattered by the individual nanoantennas. The restored alternating phase of the individual nanoantennas is clearly demonstrated by the color-coded image in Figure 6c. The restored phase image was further processed to demonstrate the high precision of the measurement. The correct phase restoration is clearly documented by the cross-section phase profile in Figure 6d taken along the dashed line in Figure 6c. Except for one of the eight nanoantennas imaged, the restored phases match well with the values $-\pi/2$ and $\pi/2$ set by the angular rotation of the nanoantennas. To determine the precision of the phase restoration in the whole nanoantenna array, histograms shown in Figure 6d were created for both alternating phases. In the histograms, 5 central pixels in each image spot of the restored nanoantenna array were processed. The bin counts representing the number of occurrences of specific phase values were fitted by the normal distribution, and the mean value $\mu$ and the standard deviation $\sigma$ were calculated. The obtained values $\mu$ = 1.49 rad, $\sigma$ = 0.12 rad and $\mu$ = $-1.50$ rad, $\sigma$ = 0.15 rad show that the precision of the measurement is maintained well below $\lambda_c/40$ even when the phase is restored in a highly sensitive measurement using signal scattered by a single nanoantenna.

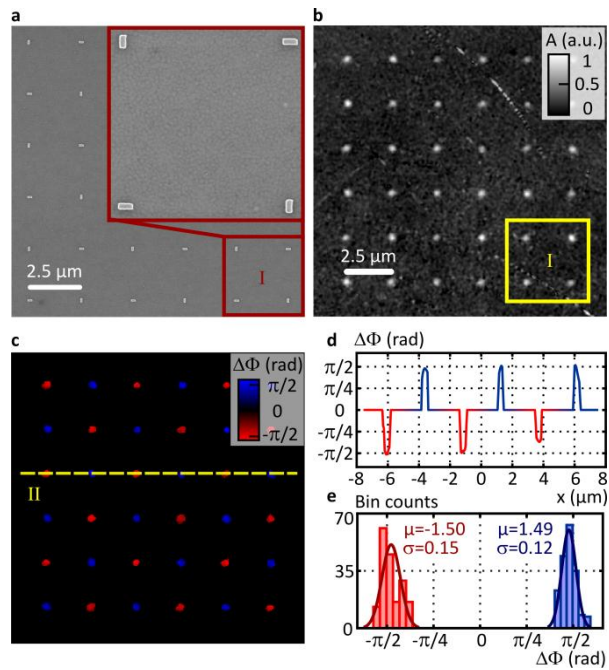

**Figure 6** Phase imaging demonstrating the sensitivity to individual nanoantennas. (a) SEM image of nanoantenna array used as a sample. (b) Amplitude of light scattered by individual nanoantennas reconstructed from the correlation records. (c) Color-coded phase altered by individual nanoantennas. (d) Cross-section phase profile along dashed line in (c). (e) Histograms of the alternating phase set by the individual nanoantennas using normal distribution fitting.

**Discussion**

Q4GOM presented in the article introduces a new concept of non-scanning imaging of plasmonic metasurfaces providing unprecedented spatial resolution and light sensitivity. Unlike recently demonstrated coherence controlled holographic microscopy,[36] in which the disturbing leakage light must be cancelled, this light is effectively utilized in Q4GOM. The leakage field provides a pure reference phase, remaining unaffected by the metasurface structure and ensuring an excellent accuracy of the measurement. The increased performance of Q4GOM is achieved using geometric-phase elements allowing selective polarization control of scattered and leakage light, referred to as polarization-division multiplexing. Unique capabilities of 4G optics elements ensure the high light sensitivity of the measurement and enable the inherently stable common-path arrangement of the experiments to be performed with broadband spatially incoherent light.

The experiments carried out with metasurfaces justify Q4GOM to be a powerful and versatile technique that can be easily deployed even in high-demand measurements. The spatial resolution approaching the diffraction limit of optical microscopy and the capability to restore phase from light scattered by a single nanoantenna open entirely new possibilities of Q4GOM in the advanced inspection of metasurfaces. The unique ability of Q4GOM to realize a complete mapping of alterations in amplitude and phase performed by individual nanoantennas, which was clearly demonstrated in experiments with vortex structures, represent a significant progress in the diagnostics of metasurfaces. The availability of highly accurate information on basic building blocks of real metasurfaces offers a major contribution to theoretical studies, design process and implementation of complex metasurface elements. Combination of high spatial resolution, measurement accuracy and single shot operation might be of great importance also for study of fundamental effects and dynamics of phase variations in active metasurfaces or plasmonic biosensors. For example, the method can be used for real-time monitoring of active metasurfaces, possible identification of failing or lagging building blocks, or revealing undesirable interaction between individual building blocks.

The basic principle of Q4GOM and pilot experiments show that potential of the method goes far beyond demonstrated metasurface measurements. The combination of polarization-division multiplexing with the self-interference of spatially incoherent light developed in Q4GOM can be directly applied to a quantitative mapping of phase differences introduced between any orthogonal polarization components. This capability is highly appreciated in optical community and opens a wide range of applications. The measurement of alterations in the geometric phase realized by a spatial change of anisotropy axis in polymer liquid crystals is one of the possible examples. Because the efficiency of these elements is extremely high and the leakage light is negligibly weak, the primary and complex conjugate waves are used to capture the correlation records. Another important application is the quantitative evaluation of the differences in the dynamic phase (i.e., the refractive index) caused by birefringence. In this case, the phase differences introduced between orthogonal linear polarizations by individual pixels of a spatial light modulator can be evaluated with high accuracy as shown in the calibration measurements. Another promising research topic is the study of spatial phase variations introduced between orthogonal circular polarizations by chiral samples exhibiting circular birefringence.

**Materials and methods**

*Sample preparation*

The substrate chosen for the sample fabrication was conductive silicon with the following layers deposited using e-beam evaporator: 3 nm Ti adhesion layer, 200 nm Au mirror layer, 3 nm Ti adhesion layer and 110 nm thick $SiO_2$ spacer layer. Substrates were covered by 125 nm thick layer of CSAR 62 e-beam resist (Allresist AR-P 6200.07, 4000 rpm) and Electra 92 conductive polymer (Allresist AR-PC-5090). Structures were exposed using a Tescan Mira microscope with RaithElphy system (30 kV, base dose 110 µC/cm$^2$). All patterns were proximity corrected using GnISys BEAMER software. The samples were developed in amyl acetate developer (Allresist AR 600-546) and rinsed in isopropyl alcohol. The fabrication was finished by deposition of 3 nm Ti adhesion layer, 30 nm Au and lift-off in Dioxolane (Allresist AR-P 600-71). SEM images were taken using FEI Verios 460L in immersion mode.

*Design and implementation of add-on imaging module*

The add-on geometric-phase imaging module is composed of the Thorlabs cage system, the GPG, the linear polarizer LP and two Nikon lenses $L_1$ and $L_2$ forming a 4f system. The GPG was custom-made by Boulder Nonlinear Systems as a polymer liquid crystal grating with the spatially modulated orientation of the anisotropy axis [size (10×10) mm$^2$, design wavelength 633nm, spectral bandwidth 100 nm, efficiency >90 %]. The steering angle between +1st and −1st diffraction order of the GPG and the magnification of the 4f system were optimally chosen to create the carrier frequency enabling spatial separation of the interfering waves in the Fourier domain. In the final design, the steering angle of 8.16$^o$ was chosen and combined with the magnification 4× of the 4f system. To maintain the proper carrier frequency in the entire field of view, the 4G imaging module was designed to minimize optical aberrations. Hence, the well corrected Nikon camera objective (f = 50 mm) and the Nikon tube lens (f = 200 mm) were used as the first and second lenses of the 4f system.

*Theoretical framework of Q4GOM*

In the performed measurements, the broadband LHCP light emitted by a spatially incoherent source was used as the driving light for metasurfaces capable of altering the Pancharatnam-Berry phase. These metasurfaces are composed of identical nanoantennas with varying rotation angles generating scattered and leakage waves. The scattered wave carries information about local variations of the geometric phase that are determined by the angular rotation of the nanoantennas and introduced while changing the polarization state from LHCP to RHCP. The leakage wave maintains LHCP of the driving light and provides a well-defined reference phase. The phase image is restored from an off-axis hologram created by the self-interference of the scattered and leakage waves having mutual angular inclination created by the GPG in the add-on imaging module. To achieve interference, the optical path difference between scattered and leakage waves may not exceed the coherence length given by the spectral width. Because the spatially incoherent source is used, the light scattered by a single nanoantenna is correlated only with the leakage light originating from a very close neighborhood. Hence, the entire hologram is generated as an incoherent superposition of point holograms emerging because of the interference of overlapping image spots created by the scattered and leakage light. The phase of the metasurface is restored from the mutual coherence function representing one of the cross terms of the hologram. To support and optimize the experiments, the simulation model of Q4GOM was developed, in which the transformation of the geometric phase by the metasurface and the GPG is described using the Jones calculus,[3] the spatial correlation of the scattered and leakage light assessed by the van Cittert-Zernike theorem, and the acquired correlation records calculated using the convolution approach. This model enables evaluation of the phase restoration in dependence on the spatial resolution of the used optical elements and the spatial correlation of light in the metasurface plane.

*Numerical simulation*

As the numerical calculation of the fields produced by the whole benchmark metasurfaces would be time-consuming and rather inefficient, we employed a procedure developed in our previous work.[36] In the first step, the field produced by a single nanoantenna was calculated using FDTD method (Lumerical FDTD Solutions). The incorporation of the mutual interaction between neighboring elements was ensured by considering a small (3 × 3) nanoantenna array surrounded by PML boundary conditions. Meanwhile, to simulate the incoherent illumination used in the experiment, only the central antenna was illuminated (TFSF source). In the second step, we determined the response of a single nanoantenna in terms of its dipole vector (extracted from the FDTD simulations), using the Green's function formalism.[49] Replacing the nanoantenna with a point dipole would be inadequate for the analysis of its near-field, but since our measurements are far-field based, this approximation is acceptable. The far-field image of the dipole was obtained by expanding the free space dyadic Green's function into propagating and evanescent waves and retaining only those that can pass through the microscope. More specifically, the integration in the Fourier space was limited to lateral wavevectors that satisfy the condition $|\mathbf{k}_\perp| < 2\pi/\lambda \mathrm{NA}$, where NA is the numerical aperture of the used microscope objective. Note that the presence of the substrate necessitates to multiply all the waves by their corresponding Fresnel coefficients. In the third and final step, the image of the whole metasurface was calculated by taking the convolution of the single nanoantenna response function with the distribution function of the nanoantennas constituting the metasurface.

*Processing and reconstruction of holographic records*

Reconstruction of holographic records is based on Fourier filtering which is well established in off-axis holography. The holographic record taken with an achromatic carrier frequency is Fourier transformed and the spatial spectrum of the true holographic image is separated and cut out using the window function. After inverse Fourier transformation raw amplitude and phase images are reconstructed. Thanks to the common-path configuration of the experimental setup, the reconstructed phase images are minimally affected by background distortion which is troublesome in traditional holographic techniques. The raw amplitude and phase images might be further processed in order to obtain unwrapped phase or segmented images. Applied reconstruction algorithm is based on the procedure described by Zikmund et al.[50]

**Acknowledgment**

This work has been supported by the Grant Agency of the Czech Republic (GA15-21581S, GA18-01396S) and by MEYS (LM2015062 Czech-BioImaging). The research was partially carried out under the project CEITEC 2020 (LQ1601) with a financial support from the Ministry of Education, Youth and Sports of the Czech Republic under the National Sustainability Programme II. Part of the work was carried out with the support of CEITEC Nano Research Infrastructure (ID LM2015041, MEYS CR, 2016–2019), CEITEC Brno University of Technology. P. Bouchal has been supported by scholarship awarded by the Czechoslovak Microscopy Society.

**Correspondence:** Petr Bouchal (petr.bouchal@ceitec.vutbr.cz)

**Author's contributions**

P.B. proposed principle of quantitative measurement, performed experiments and processed the experimental data. P.D. assisted with experiments and discussed the results. F.L. designed and J.B. and A.F. prepared the samples. M.H. and V.K. performed the theoretical simulations of metasurfaces. Z.B. contributed with theoretical model and numerical simulations of the measurement. P.B., Z.B. and R.C. designed polarization imaging module used in experiments. P.B., P.D., F.L. and Z.B. wrote the manuscript. S.L. and T.Š. directed and supervised the project. All authors discussed the results and commented on the manuscript.

**REFERENCES**

1   Yu N, Capasso F. Flat optics with designer metasurfaces. *Nat Mater* 2014; **13**: 139–150.
2   Kildishev A V., Boltasseva A, Shalaev VM. Planar Photonics with Metasurfaces. *Science (80- )* 2013; **339**: 1232009–1232009.
3   Hsiao H-H, Chu CH, Tsai DP. Fundamentals and Applications of Metasurfaces. *Small Methods* 2017; **1**: 1600064.
4   Neshev D, Aharonovich I. Optical metasurfaces: new generation building blocks for multi-functional optics. *Light Sci Appl* 2018; **7**: 58.
5   Lin D, Fan P, Hasman E, Brongersma ML. Dielectric gradient metasurface optical elements. *Science (80- )* 2014; **345**: 298–302.
6   Khorasaninejad M, Chen WT, Devlin RC, Oh J, Zhu AY, Capasso F. Metalenses at visible wavelengths: Diffraction-limited focusing and subwavelength resolution imaging. *Science (80- )* 2016; **352**: 1190 LP-1194.
7   Ee HS, Agarwal R. Tunable Metasurface and Flat Optical Zoom Lens on a Stretchable Substrate. *Nano Lett* 2016; **16**: 2818–2823.
8   Fan Q, Huo P, Wang D, Liang Y, Yan F, Xu T. Visible light focusing flat lenses based on hybrid dielectric-metal metasurface reflector-arrays. *Sci Rep* 2017; **7**: 1–9.
9   Zhu Y, Li Z, Hao Z, DiMarco C, Maturavongsadit P, Hao Y *et al.* Optical conductivity-based ultrasensitive mid-infrared biosensing on a hybrid metasurface. *Light Sci Appl* 2018; **7**: 67.
10  Lee Y, Kim S-J, Park H, Lee B. Metamaterials and Metasurfaces for Sensor Applications. *Sensors* 2017; **17**: 1726.
11  Lin D, Melli M, Poliakov E, Hilaire PS, Dhuey S, Peroz C *et al.* Optical metasurfaces for high angle steering at visible wavelengths. *Sci Rep* 2017; **7**: 1–8.
12  Deng Z-L, Deng J, Zhuang X, Wang S, Shi T, Wang GP *et al.* Facile metagrating holograms with broadband and extreme angle tolerance. *Light Sci Appl* 2018; **7**: 78.
13  Chen WT, Khorasaninejad M, Zhu AY, Oh J, Devlin RC, Zaidi A *et al.* Generation of wavelength-independent subwavelength Bessel beams using metasurfaces. *Light Sci Appl* 2017; **6**: e16259.
14  Huang L, Song X, Reineke B, Li T, Li X, Liu J *et al.* Volumetric Generation of Optical Vortices with Metasurfaces. *ACS Photonics* 2017; **4**: 338–346.
15  Zhang Y, Yang X, Gao J. Twisting phase and intensity of light with plasmonic metasurfaces. *Sci Rep* 2018; **8**: 1–9.
16  Ma X, Pu M, Li X, Huang C, Wang Y, Pan W *et al.* A planar chiral meta-surface for optical vortex generation and focusing. *Sci Rep* 2015; **5**: 1–7.
17  Huang L, Chen X, Mühlenbernd H, Zhang H, Chen S, Bai B *et al.* Three-dimensional optical holography using a plasmonic metasurface. *Nat Commun* 2013; **4**: 2808.
18  Larouche S, Tsai Y-J, Tyler T, Jokerst NM, Smith DR. Infrared metamaterial phase holograms. *Nat Mater* 2012; **11**: 450–454.
19  Chen WT, Yang K-Y, Wang C-M, Huang Y-W, Sun G, Chiang I-D *et al.* High-Efficiency Broadband Meta-Hologram with Polarization-Controlled Dual Images. *Nano Lett* 2014; **14**: 225–230.
20  Ni X, Kildishev A V, Shalaev VM. Metasurface holograms for visible light. *Nat Commun* 2013; **4**: 2807.
21  Yao Y, Shankar R, Kats MA, Song Y, Kong J, Loncar M *et al.* Electrically tunable metasurface perfect absorbers for ultrathin mid-infrared optical modulators. *Nano Lett* 2014; **14**: 6526–6532.
22  Wang Q, Rogers ETF, Gholipour B, Wang C-M, Yuan G, Teng J *et al.* Optically reconfigurable metasurfaces and photonic devices based on phase change materials. *Nat Photonics* 2016; **10**: 60–65.
23  Papaioannou M, Plum E, Rogers ET, Zheludev NI. All-optical dynamic focusing of light via coherent absorption in a plasmonic metasurface. *Light Sci Appl* 2018; **7**: 17157.
24  Li G, Chen S, Pholchai N, Reineke B, Wong PWH, Pun EYB *et al.* Continuous control of the nonlinearity phase for harmonic generations. *Nat Mater* 2015; **14**: 607–612.
25  Almeida E, Bitton O, Prior Y. Nonlinear metamaterials for holography. *Nat Commun* 2016; **7**: 12533.
26  Neuman T, Alonso-González P, Garcia-Etxarri A, Schnell M, Hillenbrand R, Aizpurua J. Mapping the near fields of plasmonic nanoantennas by scattering-type scanning near-field optical microscopy. *Laser Photon Rev* 2015; **9**: 637–649.
27  Bohn BJ, Schnell M, Kats MA, Aieta F, Hillenbrand R, Capasso F. Near-Field Imaging of Phased Array

Metasurfaces. *Nano Lett* 2015; **15**: 3851–3858.

28   Drachev VP, Cai W, Chettiar U, Yuan H-K, Sarychev AK, Kildishev A V *et al.* Experimental verification of an optical negative-index material. *Laser Phys Lett* 2006; **3**: 49–55.

29   Börzsönyi A, Kovács AP, Görbe M, Osvay K. Advances and limitations of phase dispersion measurement by spectrally and spatially resolved interferometry. *Opt Commun* 2008; **281**: 3051–3061.

30   O'Brien K, Lanzillotti-Kimura ND, Suchowski H, Kante B, Park Y, Yin X *et al.* Reflective interferometry for optical metamaterial phase measurements. *Opt Lett* 2012; **37**: 4089.

31   Gennaro SD, Sonnefraud Y, Verellen N, Van Dorpe P, Moshchalkov V V, Maier S a *et al.* Spectral interferometric microscopy reveals absorption by individual optical nanoantennas from extinction phase. *Nat Commun* 2014; **5**: 3748.

32   Lepetit L, Chériaux G, Joffre M. Linear techniques of phase measurement by femtosecond spectral interferometry for applications in spectroscopy. *J Opt Soc Am B* 1995; **12**: 2467.

33   Meshulach D, Yelin D, Silberberg Y. Real-time spatial–spectral interference measurements of ultrashort optical pulses. *J Opt Soc Am B* 1997; **14**: 2095.

34   Dolling G, Enkrich C, Wegener M, Soukoulis CM, Linden S. Simultaneous negative phase and group velocity of light in a metamaterial. *Science (80- )* 2006; **312**: 892–894.

35   Onishi S, Matsuishi K, Oi J, Harada T, Kusaba M, Hirosawa K *et al.* Spatiotemporal control of femtosecond plasmon using plasmon response functions measured by near-field scanning optical microscopy (NSOM). *Opt Express* 2013; **21**: 26631–26641.

36   Babocký J, Křížová A, Štrbková L, Kejík L, Ligmajer F, Hrtoň M *et al.* Quantitative 3D Phase Imaging of Plasmonic Metasurfaces. *ACS Photonics* 2017; **4**: 1389–1397.

37   Kim J, Li Y, Miskiewicz MN, Oh C, Kudenov MW, Escuti MJ. Fabrication of ideal geometric-phase holograms with arbitrary wavefronts. *Optica* 2015; **2**: 958–964.

38   Escuti MJ, Kim J, Kudenov MW. Geometric-Phase Holograms. *Opt Photonics News* 2016; **27**: 22–29.

39   Lee Y-H, Tan G, Zhan T, Weng Y, Liu G, Gou F *et al.* Recent progress in Pancharatnam–Berry phase optical elements and the applications for virtual/augmented realities. *Opt Data Process Storage* 2017; **3**: 79–88.

40   Allen L, Beijersbergen WM, Spreeuw RJC, Woerdman JP. Orbital angular momentum of light and the transformation of Laguerre-Gaussian laser modes. *Phys Rev A* 1992; **45**: 8185–8189.

41   Karimi E, Schulz S a, De Leon I, Qassim H, Upham J, Boyd RW. Generating optical orbital angular momentum at visible wavelengths using a plasmonic metasurface. *Light Sci Appl* 2014; **3**: e167.

42   Wang Y, Fang X, Kuang Z, Wang H, Wei D, Liang Y *et al.* On-chip generation of broadband high-order Laguerre-Gaussian modes in a metasurface. *Opt Lett* 2017; **42**: 2463–2466.

43   Faßbender A, Babocký J, Dvořák P, Křápek V, Linden S. Invited Article: Direct phase mapping of broadband Laguerre-Gaussian metasurfaces. *APL Photonics* 2018; **3**: 110803.

44   Nelayah J, Kociak M, Stéphan O, García de Abajo FJ, Tencé M, Henrard L *et al.* Mapping surface plasmons on a single metallic nanoparticle. *Nat Phys* 2007; **3**: 348–353.

45   Krenn JR, Dereux A, Weeber JC, Bourillot E, Lacroute Y, Goudonnet JP *et al.* Squeezing the Optical Near-Field Zone by Plasmon Coupling of Metallic Nanoparticles. *Phys Rev Lett* 1999; **82**: 2590–2593.

46   Bouchal P, Čelechovský R, Bouchal Z. Polarization sensitive phase-shifting Mirau interferometry using a liquid crystal variable retarder. *Opt Lett* 2015; **40**: 4567–4570.

47   Bouchal P, Chmelík R, Bouchal Z. Dual-polarization interference microscopy for advanced quantification of phase associated with the image field. *Opt Lett* 2018; **43**: 427–430.

48   Minovich AE, Peter M, Bleckmann F, Becker M, Linden S, Zayats A V. Reflective Metasurfaces for Incoherent Light To Bring Computer Graphics Tricks to Optical Systems. *Nano Lett* 2017; **17**: 4189–4193.

49   Novotny L, Hecht B. *Principles of Nano-Optics*. 2nd ed. Cambridge University Press, 2012.

50   Zikmund T, Kvasnica L, Týč M, Křížová A, Čolláková J, Chmelík R. Sequential processing of quantitative phase images for the study of cell behaviour in real-time digital holographic microscopy. *J Microsc* 2014; **256**: 117–125.